\documentclass[letterpaper,12pt,preprint]{aastex}

\shorttitle{Structure of Disks with Grain Growth}
\shortauthors{Aikawa \& Nomura}

\begin{document}


\title{Physical and Chemical Structure of Protoplanetary Disks
with Grain Growth}


\author{Y. Aikawa and H. Nomura}
\affil{Department of Earth and Planetary Sciences, Kobe University,
Kobe 657-8501, Japan}

%\and
%
%\author{R. J. Hanisch\altaffilmark{5}}
%\affil{Space Telescope Science Institute, Baltimore, MD 21218}


%\altaffiltext{5}{Patron, Alonso's Bar and Grill}


\begin{abstract}
We calculate the physical structure of protoplanetary disks by
evaluating the gas density and temperature self-consistently and
solving separately for the dust temperature. The effect of grain
growth is taken into account by assuming a power-law size 
distribution and varying the maximum radius of grains $a_{\rm max}$.
In our fiducial model with $a_{\rm max}=10\mu$m, the gas is warmer than
the dust in the surface layer of the disk, while the gas and dust have the same
temperature in deeper layers. In the models with larger
$a_{\rm max}$, the gas temperature in the surface layer is lower than in
the fiducial model
because of reduced photo-electric heating rates from small grains,
while the deeper penetration of stellar radiation warms the gas at intermediate
height.
A detailed chemical reaction network is solved at outer radii ($r \ge 50$ AU).
Vertical distributions of some molecular species at different radii are
similar, when plotted as a function of hydrogen column density $\Sigma_{\rm H}$
from the disk surface. Consequently, molecular column densities do not much
depend on disk radius. In the models with larger $a_{\rm max}$, the lower
temperature in the surface layer makes the geometrical thickness of the
disk smaller, and the gaseous molecules are confined to smaller heights.
However, if we plot the vertical distributions of molecules as a function of
$\Sigma_{\rm H}$, they do not significantly depend on $a_{\rm max}$.
The dependence of the molecular column densities on $a_{\rm max}$ is not
significant, either. Notable exceptions are HCO$^+$, H$_3^+$ and H$_2$D$^+$,
which have smaller column densities in the models with larger $a_{\rm max}$.

\end{abstract}




\section{Introduction}

Planetary systems are thought to be formed in circumstellar disks, called protoplanetary disks, around young low-mass stars. The physical
structure and evolution of the disks have been investigated by observation and
by theoretical modeling. The disks are in hydrostatic equilibrium in the vertical
direction, while the temperature is determined mainly by accretion at inner
radii ($\lesssim$ several AU) and by irradiation from the central star
at outer radii. Hence the calculation of temperature is important in determining
the physical structure of disks. In recent years several groups have been
constructing disk models by solving the equations of radiation transfer.
\citet{dal98} obtained the vertical structure
by solving one-dimensional radiation transfer at each disk radius, while
\citet{nom02} and \citet{dul02} solved the disk structure using a
two-dimensional
radiation transfer code. Although these models differ quantitatively, they 
all flare up significantly --- the scale height is larger at larger radii.
It should be noted that the density and temperature distributions
are coupled; warm temperature in the outer radii makes the flared-up
structure, which enhances the local irradiation (i.e. heating) from the
central star.
\citet{dal99} pointed out that these model disks are geometrically
too thick. In order to reproduce the observed spectral energy
distribution and images of edge-on disks, some fraction of grains should be
larger than interstellar dust \citep{dal01}. This grain growth
decreases
the vertical height of the disk photosphere where the optical depth to the
stellar radiation is unity, and thus the local heating rate by irradiation
and hence the scale height.
The presence of mm-sized grain is also required to account for the relatively
high intensity of the millimeter dust continuum emission.

The theoretical models mentioned above, however, assumed that the gas
temperature is equal to the dust temperature. Since the vertical density
distribution is determined by gas-pressure gradient, the validity of this
assumption should be
checked. Recently \citet{kd04} tackled this problem; thermal
balance is solved to obtain gas temperature, while the density distribution and
the dust temperature are fixed to those of \citet{dzn02}, in which
astronomical silicate grains of 0.1 $\mu$m were assumed. It is found
that dust and gas temperatures are equal to within 10\% for the visual
extinction of $A_{\rm v} \gtrsim
0.1$ mag, and that the gas temperature exceeds the dust temperature at larger
heights. A similar result is obtained by \citet{jfz04}, who
calculated the gas temperature by using the density distribution and the dust
temperature of \citet{dal99}. \citet{jfz04} also
investigated effect of dust settling; the gas temperature is calculated
with the dust abundance reduced by two orders of magnitude above certain
heights. They found that the dust settling increases the amount of hot
gas if PAHs remain well-mixed with the gas.

In the present paper, we calculate gas and dust temperature separately
together with the gas density distribution in the disk.
The gas temperature is obtained by assuming a simple chemical equilibrium and
detailed thermal balance,
including gas-dust heat exchange, at each position in the disk. Then the
vertical density distribution is determined from hydrostatic equilibrium.
The dust temperature is obtained
by solving the two-dimensional radiation transfer \citep{nom02}. We calculate
the temperature and density profiles self-consistently by iteratively
solving the equations.
The effect of grain growth is investigated by assuming a power-law size
distribution of grains and varying the maximum grain radius $a_{\rm max}$
\citep{mn93}.

We also calculate molecular distributions in the disk models obtained above by
solving a detailed chemical reaction network. Several molecular lines, such as
CO, HCO$^+$, HCN and CN are detected by radio observations \citep{dut97, qi03,
thi04}, which are sensitive to the region outside
$\sim 100$ AU because of the beam size ($\gtrsim 1''$) of the current
instruments. Although emission bands of CO and H$_2$O in inner hot region
($r\lesssim 1$ AU) are detected in the infrared wavelength \citep{ncm03,ctn04},
little constraints are given yet on the molecular abundance and
their spatial distribution.
Therefore we constrain our discussion of the detailed chemistry to radii $\ge 50$ AU.

\citet{aik02} and \citet{zah03} calculated the molecular
distribution using the disk model of \citet{dal99}, in which
interstellar-sized grains are well mixed with gas. The disk can be
schematically divided into three layers. In the surface layer, molecules are
dissociated by ultraviolet (UV) radiation and X-rays from the central star
and the interstellar
field, while heavy-element species are frozen onto grains in the midplane
with high density and low temperature. Gaseous molecules observed at radio
wavelengths are mostly present in the intermediate layer where the gas
temperature and density are relatively high
and the UV radiation is not too harsh. The model explains the
observational results; the freeze-out in the midplane lowers the averaged
abundances of heavy-element species and the UV radiation causes the
high abundances of radical species.
However, there are some quantitative disagreements.
The models of \citet{aik02} overestimate CO column density
in DM Tau, and underestimate column densities of organic species in LkCa15.
\citet{zah03} somewhat improved the agreement by including
scattering of stellar UV, which enhances the UV intensity and hence the
photo-dissociation rate of CO within the disk. The dissociation of CO produces
carbon atoms, which are transformed to other organic species.
Therefore it is interesting to see how the molecular abundances are affected
by grain growth, which will enhance the UV penetration into the disk.

The remainder of the paper is organized as follows.  A description
of the physical and chemical model is given in \S 2. Numerical results
of the physical structure and the molecular distribution are presented in
\S 3. In \S 4, we check the self-consistency of our model and discuss implications
for molecular line observations. A summary is given in \S 5. 

\section{Models}
\subsection{Physical Model}
In this subsection we briefly describe how we obtain the physical structure
of a disk; more detailed explanations are given in \citet{nm05}.
We consider an axisymmetric Keplerian disk around a typical
T Tauri star with mass $M_{\ast}=0.5 M_{\odot}$, radius
$R_{\ast}=2R_{\odot}$, and temperature $T_{\ast}=4000$ K
\citep[e.g.][]{kh95}.
 
The disk is in hydrostatic equilibrium in the vertical direction.
The surface density distribution, $\Sigma(r)$, is determined by assuming
a steady accretion rate \citep[e.g.][]{lp74}; the thermal
heating via viscous dissipation is equal to the gravitational energy of
accreting mass at each radius. We assume constant mass accretion rate of
$\dot{M}=10^{-8} M_{\odot}$ yr$^{-1}$ and the viscous parameter of
$\alpha= 0.01$.

\begin{figure}
\plotone{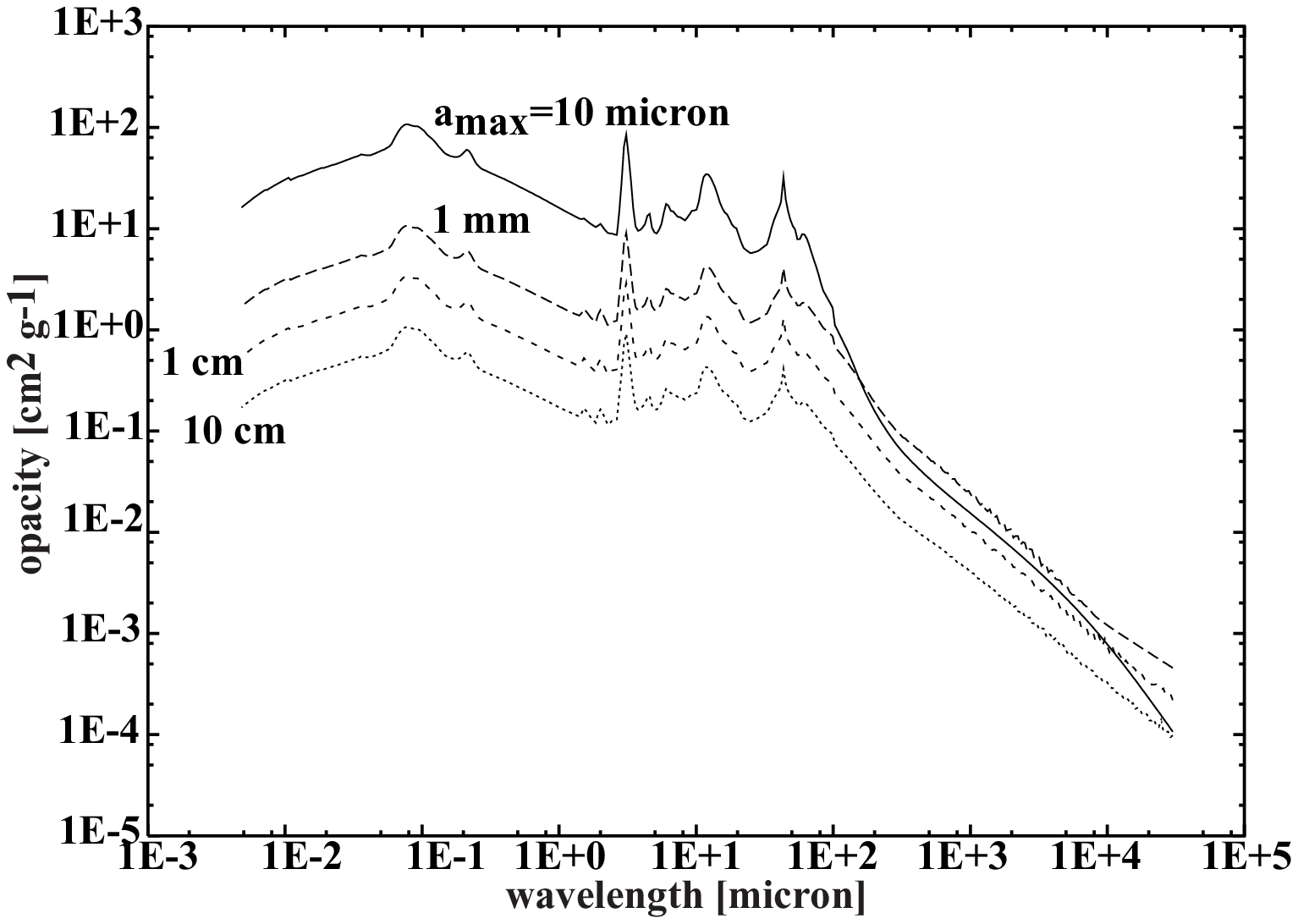}
\caption{Dust opacities in the models with $a_{\rm max}=10 \mu$m, 1 mm, 1cm,
and 10 cm.
\label{opacity}}
\end{figure}

The dust temperature is calculated assuming the local thermodynamic equilibrium and
the local radiative equilibrium between absorption and reemission by dust
grains at each position ($R, \Theta$) in the disk, where $R$ is the distance
from the central star and $\Theta$ is the angle from the rotation axis. The
equation is
\begin{equation}
\int^{\infty}_0 d\nu \kappa_{\nu}(R, \Theta)\oint d\mu d\phi
I_{\nu}(R, \Theta; \mu, \phi)=
4\pi\int^{\infty}_0 d\nu \kappa_{\nu}(R, \Theta) B_{\nu}[T_{\rm dust}(R, \Theta)], \\
\end{equation}
where $\kappa_{\nu}$, $I_{\nu}$, and $B_{\nu}(T_{\rm dust})$ represents the monochromatic
opacity, the specific intensity, and the Planck function for blackbody radiation
at a frequency $\nu$, respectively.
The energy exchange between gas and dust (Eq. [3]) is not included
in the calculation of dust temperature because it is negligible
compared with the radiative heating \citep[e.g.][]{cg97}. 
The specific intensity is calculated by solving the axisymmetric
two-dimensional radiative transfer equation,
\begin{equation}
I_{\nu}(R, \Theta; \mu, \phi) = \int^s_0 \kappa_{\nu}(R',\Theta')
\rho(R',\Theta') B_{\nu}[T_{\rm dust}(R',\Theta')] e^{-\tau_{\nu}(R',\Theta')}ds',
\end{equation}
where $\tau_{\nu} (R', \Theta')$ is the specific optical depth from a point
$(R', \Theta')$ to $(R, \Theta)$ by means of the short characteristic method
in spherical coordinates \citep{dul00}. 
As a dust model, we assume a power-law size distribution of grains, $dn/da
\propto a^{-3.5}$, where $a$ is the dust radius, and vary the maximum
grain radius $a_{\rm max}$ \citep{mn93}. The minimum grain
radius $a_{\rm min}$ is set to be 50 \AA ~ for all models.
The dust grains are assumed to be well-mixed with the gas, and the mass
ratio of the dust to the gas is fixed throughout the disk.
The dust opacity for each model is calculated using Mie theory and
plotted in Figure \ref{opacity}, 
where silicate, carboneous, and water ice are adopted as dust components.
Heating sources of dust grains are the viscous dissipation at the
midplane of the disk and the irradiation from the central star.

The gas temperature is determined by detailed energy balance at each position
in the disk.
In the surface layer of the disk, photo-electric heating is the dominant
heating mechanism. For our model with $a_{\rm max} = 10 \mu$m, we adopt the
photo-electric heating rate of \citet{wd01b}
($R_{\rm v}=5.5, b_{\rm c}=3\times 10^{-5}$, Case B), in which the
carboneous grains have PAH-like properties in small size limit and
graphite-like properties at larger sizes,
 and the maximum size of carbonaceous grain is $\sim 10 \mu$m. 
%REPEATED
In the models with different
$a_{\rm max}$, the photo-electric heating rate is calculated to be proportional
to the total number of grain particles, since small grains dominate in both
the photo-electric heating rates and the particle number.

Radiative cooling by line transitions among fine-structure levels
of OI (63 $\mu$m) and CII (158 $\mu$m), and by rotational line
transitions of CO is taken into account, and dominates the cooling
process in the very surface layer of the disk. The level populations of
OI and CII are 
obtained by solving the equation of statistical equilibrium, while the level
populations of CO are assumed to be in LTE. The cooling rate is calculated 
using the 
photon escape probability method \citep[e.g.][]{djo80}, while the abundances 
of the coolants
(OI, CII, and CO) are calculated using the chemical equilibrium equation given 
by \citet{nl97}.
%The equation requires the abundance of hydrogen
%molecule, which is determined by the formation on grain surfaces and
%photodissociation. We assume the H$_2$ formation rate of $7.5 \times 10^{-18}
%T^{1/2} n_{\rm H} n$(H) cm$^3$ s$^{-1}$  ({\sc reference}). Level population
%of H$_2$ and line shielding (Federman et al. 1979) are calculated to
%estimate the photo-dissociation rate.

In the dense region of the disk the energy exchange between gas and dust through
collisions is an important heating or cooling process. We adopt the rate for
the energy exchange
\begin{equation}
\Lambda_{\rm gr}=3.5 \times 10^{-34} n_{\rm H}^2 T_{\rm gas}^{0.5} (T_{\rm gas}-
T_{\rm dust}) ~ {\rm erg} ~ {\rm cm}^{-3} ~ {\rm s}^{-1},
\end{equation}
following \citet{bh83} and \citet{th85}.
In the models with grain growth, the rate is reduced in proportion to
the total cross section of grain particles.

The heating by X-ray irradiation from the central star and viscous
dissipation within the disk are not included in our model for simplicity.
The viscous heating is more important than the photoelectric heating,
if the phenomenological viscous parameter $\alpha$ is larger than a critical
value $\alpha_{\rm cr}$. In the case of $a_{\rm max}=10 \mu$m,
$\alpha_{\rm cr}$ is 0.05 for $r = 1$ AU and 0.1 for $r > 50$ AU, while
in the case of $a_{\rm max}
=10$ cm, $\alpha_{\rm cr}$ is $\sim 0.01$. 
%Even if the alpha is larger than
%$\alpha_{\rm cr}$, photoelectric heating dominates in the very surface
%region of the disk, in which the gas temperature deviates from the dust
%temperature. 
The X-rays could be the dominant heating source at the
surface layer, if the X-ray
luminosity is as high as 1$\times 10^{29}$, 5$\times 10^{29}$ and 1$\times
10^{30}$ ergs s$^{-1}$ for $r=50$AU, $r=100$AU, and $r>200$AU,
respectively, in the case of $a_{\rm max}=10 \mu$m
%[PUT VALUE HERE] 
\citep[e.g.][]{gni04,gh04}.
We will take X-ray heating into account in future work.

%\subsubsection{Ultraviolet radiation fields}
Protoplanetary disks are irradiated by UV radiation from the central star and
the interstellar field which affects the gas temperature and chemistry through
photo-electric heating, photodissociation and photoionization. Many
classical T Tauri stars have excess UV radiation compared to the main-sequence
stars with similar effective temperatures \citep[e.g.][]{hg86}.
In the present paper we adopt the stellar UV radiation
observed towards TW Hya \citep{cos00}; black body radiation
with $T_{\ast}=4000$ K from the star and thermal bremsstrahlung emission
with $T_{\rm br}=2.5\times 10^4$ K and $n_e^2 V= 3.68 \times 10^{56}$
cm$^{-3}$,
where $n_{\rm e}$ and $V$ are the electron number density and the volume of the
emission region. Contribution of Ly $\alpha$ line emission is
not taken into account in the present work.
% for generalizing our model as TW Hya has strong Ly $\alpha$ line emission
%compared with other T Tauri stars.
The interstellar UV field, on the other hand, is given by
Draine field \citep{dra78} for the wavelength range $912$ \AA $< \lambda < 2000$
\AA ~ and \citet{vdb82} for $\lambda > 2000$ \AA.

Transfer of UV radiation is calculated 1+1 dimensionally; in cylindrical
coordinates
\begin{equation}
F_{\nu}(r, z)=F^{\rm ISRF}_{\nu} e^{-\tau_{\nu, z}} +
 \int_0^{\tau_{\nu, z}} S_{\nu}(\tau_{\nu} ')e^{-(\tau_{\nu, z}-\tau_{\nu} ')} d\tau_{\nu} ', \label{eq:UVtrans}
\end{equation}
where $\tau_{\nu, z}$ represents the optical depth integrated vertically 
from the disk surface to the height $z$. The first term represents the
(attenuated) interstellar UV radiation.
%Since disk is not hot enough to emit UV photons,
The second term represents the contribution of the stellar UV radiation
which is scattered on dust grains towards the disk midplane:
\begin{equation}
S_{\nu}(r, z)= \omega_{\nu} \frac{\pi^2}{2}\frac{R_{\ast}^2}{R^2} F^{\rm star}_{\nu}e^{-\tau_{\nu, \ast}},
\end{equation}
%{\sc check the equation in the notebook}
where $\omega_{\nu}$ and $\tau_{\nu, \ast}$ represent the monochromatic
albedo and the specific optical depth from the central star, respectively,
and $R^2=r^2+z^2$ is a square of the distance from the central star.

\subsection{Chemical Model}
The abundance and distribution of molecules are obtained by solving the 
molecular evolution equations 
at each position in the disk. The basic equation can be written as
\begin{equation}
\frac{dn(i)}{dt}=\Sigma_j \alpha_{ij} n(j) + \Sigma_{j, k}\beta_{ijk}n(j)n(k),
\end{equation}
where $\alpha_{ij}$ and $\beta_{ijk}$ are reaction rate coefficients
and we adopt as initial conditions, molecular cloud abundances.
Chemical equilibrium is reached in a relatively short time scale $\lesssim
10^4$ yr in the surface PDR (i.e. photon-dominated region) and in the
freeze-out layer in the midplane.
Although chemistry in the intermediate layer shows more complex behavior,
it is almost in a pseudo-steady state at $t=10^5-10^6$ yr, and the molecular
column densities do not change much during this period. Hence we will present
abundance distributions at $1\times 10^6$ yr in the following sections.

We do not include hydrodynamic motions (turbulence and accretion) of disk
material and time-dependent grain growth (i.e. grain radius is fixed).
Coupling these effects with a detailed chemical reaction network
could be important, but it is very time-consuming \citep{ilg04}.
So far, there are many studies which do not include hydrodynamics 
and yet reproduce the observed molecular abundances and improve our
understanding of disk chemistry \citep[e.g.][]{ah99,wl00,mim02,kd04,jfz04,
sem05}. The primary aim of the present work is to investigate how grain
growth changes the distribution of physical parameters
(density, temperature and UV radiation) and molecular abundances in
a pseudo-steady state. The effect of hydrodynamic motions will be briefly
discussed in \S 4.2.

Calculation of molecular abundances at a fixed
position may also appear inconsistent with our derivation of physical
structure, in which we assumed steady accretion. It should be noted
that the assumption of steady accretion is used to determine the surface
density $\Sigma(r)$ and temperature. Gravitational energy released by
the accretion is much less than the irradiation from the central star at
radii $r\gtrsim$ several AU, in the region in which we calculate 
molecular abundances.
Hence our model of the outer disk is almost identical to a passive disk
with a given surface density. In addition, our results show that the vertical
distributions of molecular abundance at different radii are similar, which
suggests that inclusion of accretion and radial mixing would not
significantly modify our results at least in the outer disk.

The set of reactions and rate coefficients are given by the
``new standard model'' (NSM) \citep{th98,osa99}.
Although the original NSM is for gas-phase chemistry,
we extend it to include deuterium chemistry, adsorption of
molecules onto grains, and thermal desorption
\citep[][and references therein]{ah99}.
The rates of adsorption and H$_2$ formation are proportional to the
total surface area of grain particles and thus are reduced in the model
with large grains. The formation rate of H$_2$ is also assumed to be proportional
to the sticking probability of H atoms, which is given by \citet{th85}.
%Since T Tauri stars are strong X-ray emitters,
%with X-ray luminosities in the range $10^{29}-10^{31}$ erg s$^{-1}$
%(Montmerle et al. 1993; Glassgold et al. 1997), we include ionization by
%X-rays and X-ray-induced photolysis in the chemical reaction network
%(Maloney, Hollenbach, \& Tielens 1996; Aikawa \& Herbst. 2001), and present
%results with and without X-rays from the central star.

The rates of photodissociation and photoionization are calculated by
\begin{equation}
k_i=\int^{\infty}_{0} \sigma_{i}(\nu) \frac{1}{h\nu} 
\int_{\Omega} I_{\nu} \, d \Omega d \nu
\label{eq: rate1}
\end{equation}
where $h$ is Planck's constant, and $I_{\nu}$ the intensity of
radiation field at frequency $\nu$. 
Some molecules can be dissociated by absorption of
continuum radiation, while others need discrete line transitions or
a combination of the two \citep{vds88}. We adopt the
cross sections $\sigma_{i}(\lambda)$ given by \citet{vds88} and
updated by \citet{jsh95} and \citet{jvb95}. For species for which no
data on the cross sections are available, a rate given by a similar
type of molecule was adopted.

The hydrogen molecule and CO are dissociated by absorption of FUV ($912-1110$ \AA)
and discrete line transitions. Because of their high abundances, self-shielding
and mutual shielding of CO by H$_2$ are important. We calculate their
photodissociation rates at each position in the disk as follows. First, the
UV flux at $912-1110$ \AA ~ is calculated by integrating the spectrum and then
dividing it by the interstellar FUV field, which gives the
enhancement factor $G$ of the photodissociation photon flux. Second,
vertical column densities of H$_2$ and CO are calculated by integrating their
number densities from the disk surface to each position.
The self-shielding factor of H$_2$ is given as a function of H$_2$ column
density following \citet{db96}, while that of CO is given
as a function of H$_2$ and CO column density
following \citet{lee96}. Finally, we multiply the interstellar
photodissociation rates of H$_2$ and CO ($4.5\times 10^{-11}$ s$^{-1}$ and
$2.0\times 10^{-10}$ s$^{-1}$, respectively) by the enhancement factor and
their shielding factors.

\section{Results}
\subsection{Physical Structure}
\subsubsection{Vertical Structure}
\begin{figure}
\plotone{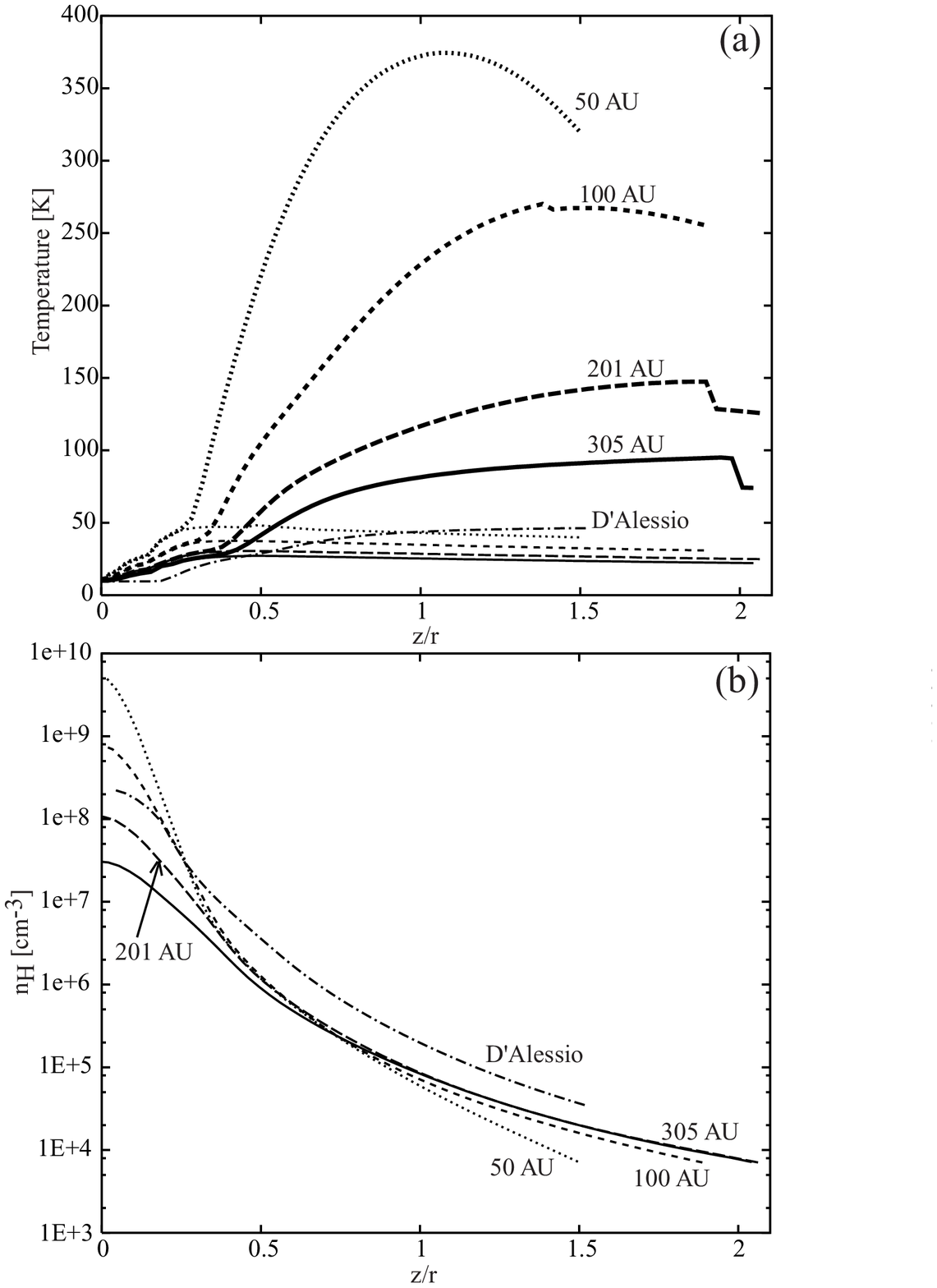}
\caption{Vertical distribution of gas (thick) and dust (thin lines)
temperature and ($b$) gas density at radii of 305 AU (solid), 201 AU
(long-dashed), 100 AU (dashed), and 50 AU (dotted lines). Dot-dashed lines
represent temperature and density distributions at $r=290$ AU in the disk
model of \citet{dal99}.
\label{rhotem_z}}
\end{figure}

Figure \ref{rhotem_z} $a$ shows the vertical distribution of gas and
dust temperatures at radii of 50, 100, 201 and 305 AU.
The maximum size of dust grain is set to be 10 $\mu$m. At
$z/r\lesssim 0.5$, gas and dust are energetically well-coupled
because of the high density. On the other hand, the gas temperature
is much higher than the dust temperature at $z/r\gtrsim 0.5$. In these surface
regions the gas temperature is determined by photo-electric heating and
line cooling of CII and OI. Sudden changes of gas temperature at
$z/r\sim 1.5-2$ appear where molecular hydrogen dominates atomic
hydrogen in collisionally exciting CII.
The difference between the gas and dust temperatures is important in predicting the
the molecular line intensity from a disk \citep{jfz04}.
The gas temperature in our model is lower than
those in \citet{kd04} and \citet{jfz04}
because of different rates of photo-electric heating on PAHs and dust
grains.

Figure \ref{rhotem_z} $b$ shows the distribution of gas density, which is
denoted by the number density of hydrogen nuclei
$n_{\rm H}(=2\times n({\rm H}_2)+n({\rm H})+n({\rm H}^+))$.
The disk is puffed up by the high gas temperature in the surface regions.
It can also be seen that the disk is flared; i.e. the disk height increases
with radius. 
%{\sc (What is the boundary condition for the disk surface ?)}

\begin{figure}
\plotone{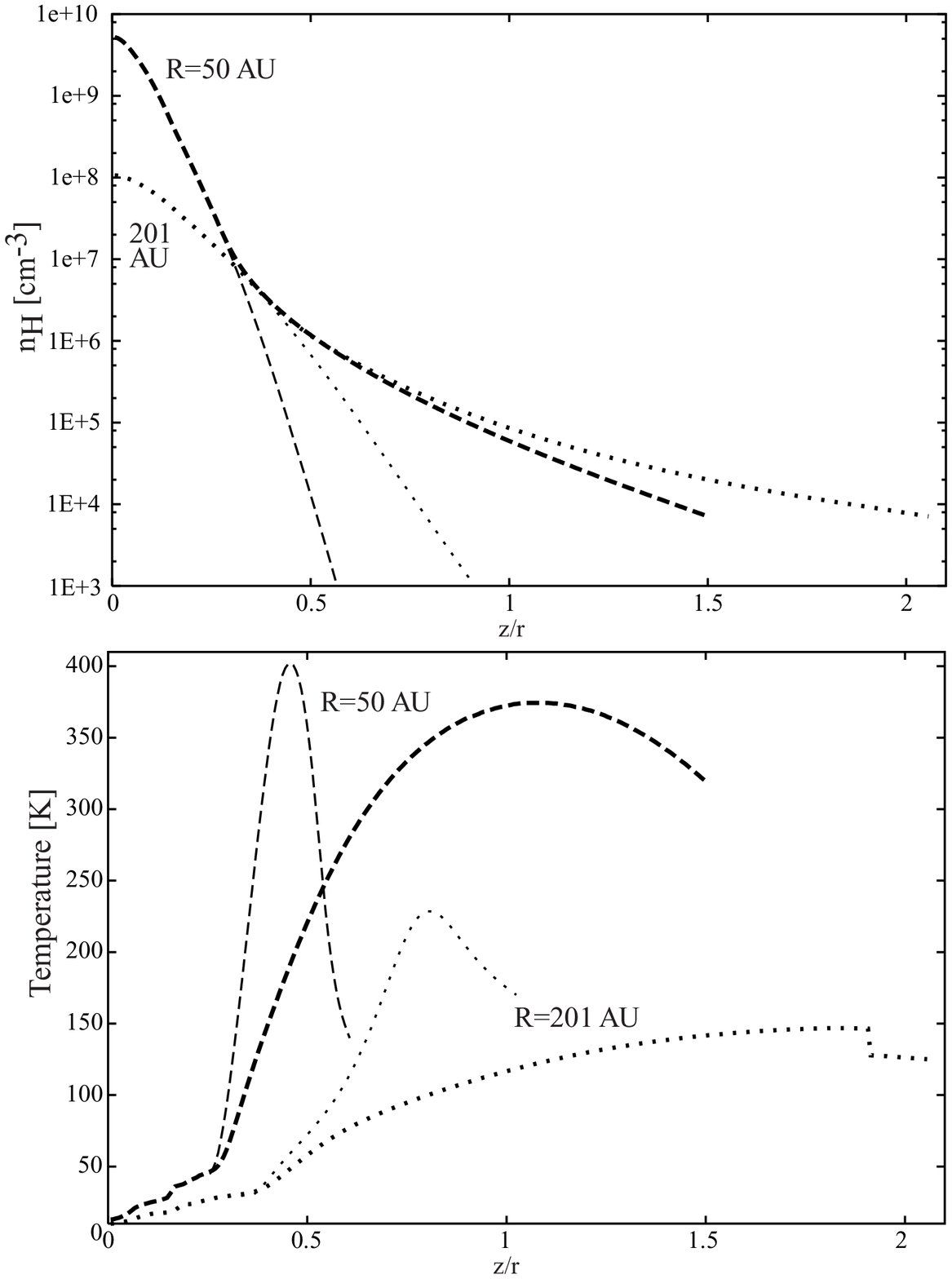}
\caption{Vertical distribution of ($a$) density and ($b$) temperature
of the gas at radii of 50 AU (dashed lines) and 201 AU (dotted lines).
Thin lines represent a model in which the gas density is determined by
assuming the gas and dust temperatures are equal. Gas temperature and density
are solved self-consistently for thick lines.
\label{cf_models}}
\end{figure}

It should be noted that the distributions of gas density, gas 
and dust temperatures are obtained self-consistently in the present work,
while \citet{kd04} and \citet{jfz04}
calculated gas temperature by adopting the gas density and the dust
temperature of \citet{dzn02} and \citet{dal99}, respectively.
For comparison with those, we have calculated the gas density by assuming the gas and
dust temperatures are equal, and then evaluated the gas temperature as in
previous studies. Figure \ref{cf_models} shows the resultant gas density (a) and
temperature (b), which are compared with our self-consistent model. It can be
seen that if the gas and dust temperatures are assumed to be equal,
the gas density is lower than the self-consistent model in the disk surface.
The self-consistent calculation of gas
density is important in the prediction of molecular line intensities and
the gas heating and cooling processes. The lower gas density at the disk
surface leads to weaker line intensity 
if the line emission mainly comes from the hot surface layer. It also
leads to an underestimate of the gas-grain collision
rate, which is proportional to square of the density.
% and dominates the gas cooling process in dense region.
The gas temperature in the disk surface is overestimated in the
inconsistent model, since the lower dust density in the inner
disk leads to an underestimate of the absorption of UV photons from
the central star, and hence to an overestimation of the photoelectric heating
rate.
On the other hand, the gas densities in the two models are similar
in the region close to the midplane. Consequently, the location of the
absorption surface where the dust optical depth to the stellar radiation
is unity is almost the same in the two models.
%The gas temperature in the disk surface is overestimated in the
%inconsistent model, since %[PUT REASON HERE]}

The dot-dashed line in Figure \ref{rhotem_z} shows the distribution of
($a$) dust temperature and ($b$) gas density
at $r=290$ AU in a disk model by \citet{dal99}.
The deviation of our dust temperature from theirs in the surface region 
is caused by a different treatment of radiative transfer;
a radial component of the flux keeps the dust temperature almost constant
at $z/r\gtrsim 0.5$ in our model, while the temperature is lower at smaller
height $z$ in the 1+1D model of \citet{dal99}.
In the deeper layers ($z/r \lesssim 0.5$), on the other hand, our model
gives a higher temperature than \citet{dal99}, because we use
the frequency dependent opacity in calculating the reprocessed radiation,
while they use mean opacity \citep[see][]{dzn02}.
Their gas density profile is similar to ours especially at larger disk
radii (except for the difference in the total column density by a factor
of $\sim 5$), since their dust temperature profile is somehow similar
to ours.

We also calculated a model with ``Dark Cloud dust'', in which the dust size
distribution 
of \citet{wd01a} ($R_{\rm v}=5.5, b_{\rm c}=3\times 10^{-5}$,
Case B) is adopted. The resultant disk structure is similar to Figure
\ref{rhotem_z} because 
%a number of small grains and thus 
the dust opacity is similar to that in the $a_{\rm max}=10 \mu$m model.

\subsubsection{Dependence on Grain Size}
\begin{figure}
\plotone{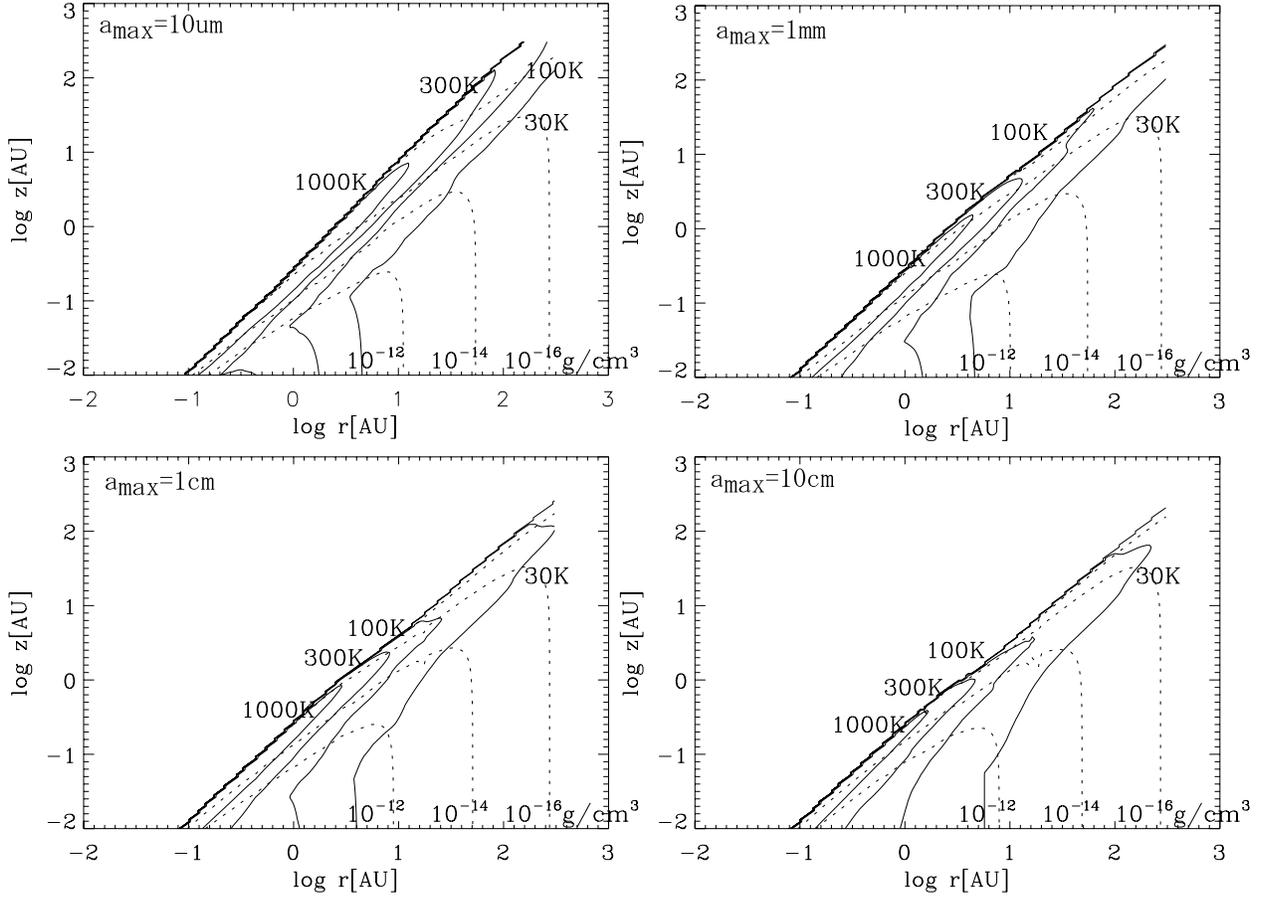}
\caption{Two-dimensional ($r,z$) contour plot of gas temperature (solid lines)
and density (dashed lines) in the disk models with $a_{\rm max}= 10 \mu$m,
1 mm, 1 cm, and 10 cm.
\label{phys_2D}}
\end{figure}

\begin{figure}
\plotone{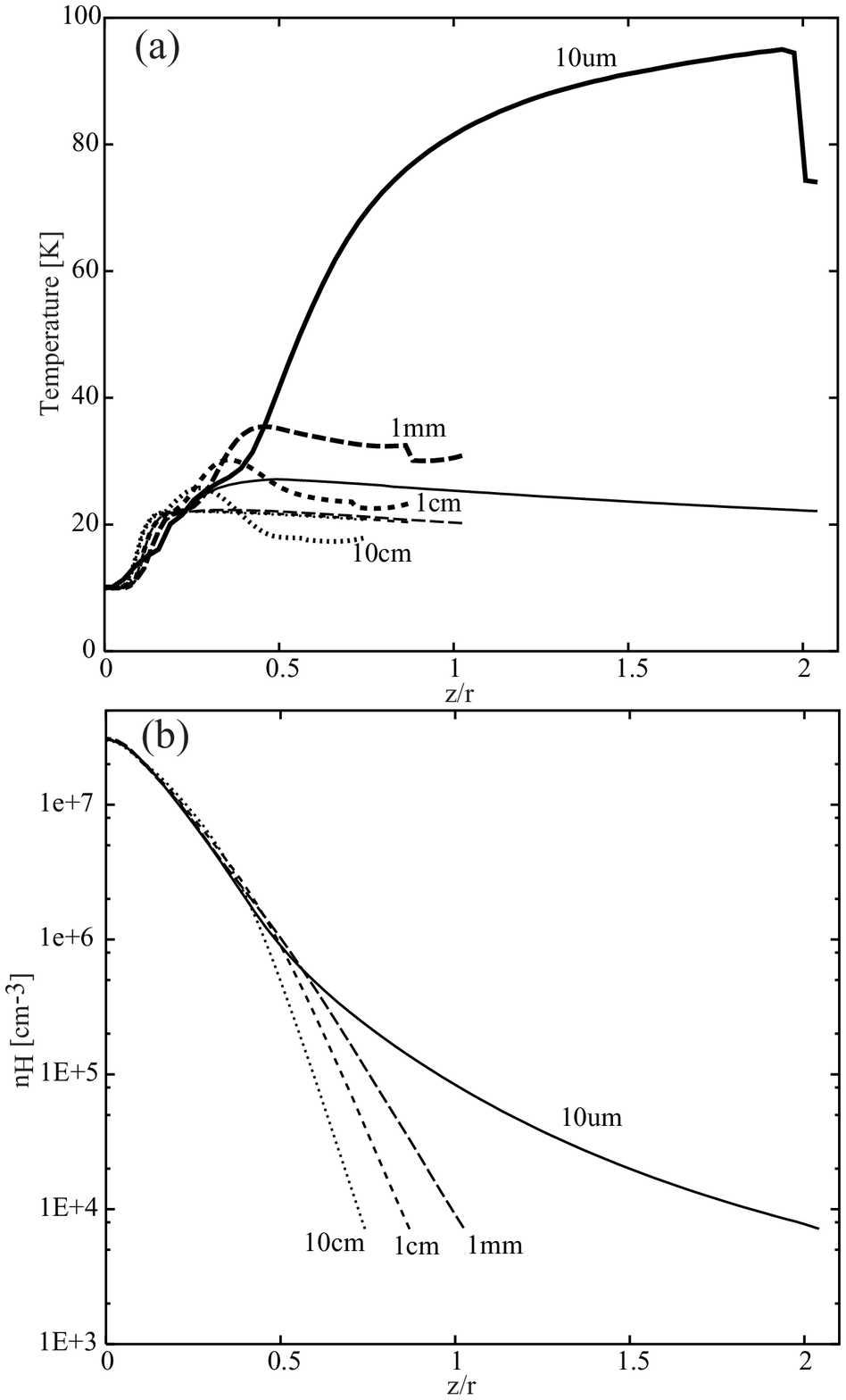}
\caption{Vertical distribution of gas (thick) and dust (thin lines)
temperature and ($b$) gas density at radius of 305 AU. The maximum radius of
dust grains is 10 $\mu$m (solid), 1 mm (long-dashed), 1 cm (dashed), and 10 cm
(dotted lines).
\label{comp_amax}}
\end{figure}

Figure \ref{phys_2D} shows two-dimensional ($r, z$) distributions of gas
temperature and density in the models with $a_{\rm max}=10 \mu$m, 1mm,
1 cm, and 10 cm. For a closer look, Figure \ref{comp_amax} $a$ shows the vertical
distributions of gas and dust temperatures at radius of 305 AU in these models.
The gas temperature in the surface region ($z/r\gtrsim 0.5$)
is lower in the models with larger $a_{\rm max}$ because of the reduced
photo-electric heating rate on small grains. On the other hand, the gas temperature
at intermediate height ($z/r\sim 0.3$) is higher in the models with larger
$a_{\rm max}$, because of deeper penetration by stellar radiation.

The model with $a_{\rm max}=10 \mu$m gives higher dust temperature
in the surface region than other models. This is caused by the dependence of dust
opacity on grain size (Figure \ref{opacity}). At the wavelength of stellar
radiation ($\sim 1 \mu$m), the dust opacity varies significantly with
$a_{\rm max}$. For example, it is larger by more than an order of magnitude
in the model with $a_{\rm max}=10\mu$m than in $a_{\rm max}=1$ mm.
At the wavelength of dust thermal emission ($\sim 100 \mu$m), on the other
hand, the dependence of dust opacity on $a_{\rm max}$ is less significant;
for example, the dust opacity is almost the same in the models with
$a_{\rm max}=10\mu$m and 1 mm.

Figure \ref{comp_amax} $b$ shows vertical distribution of gas density
at radius of 305 AU for models with various $a_{\rm max}$. The disks
with smaller $a_{\rm max}$ are more puffed up because of the higher gas
temperature in the surface layer.

%comparison with fig 6 of Jonkheid ?

\subsection{Molecular Abundance}
\subsubsection{Vertical and Radial Distribution}
\begin{figure}
\plotone{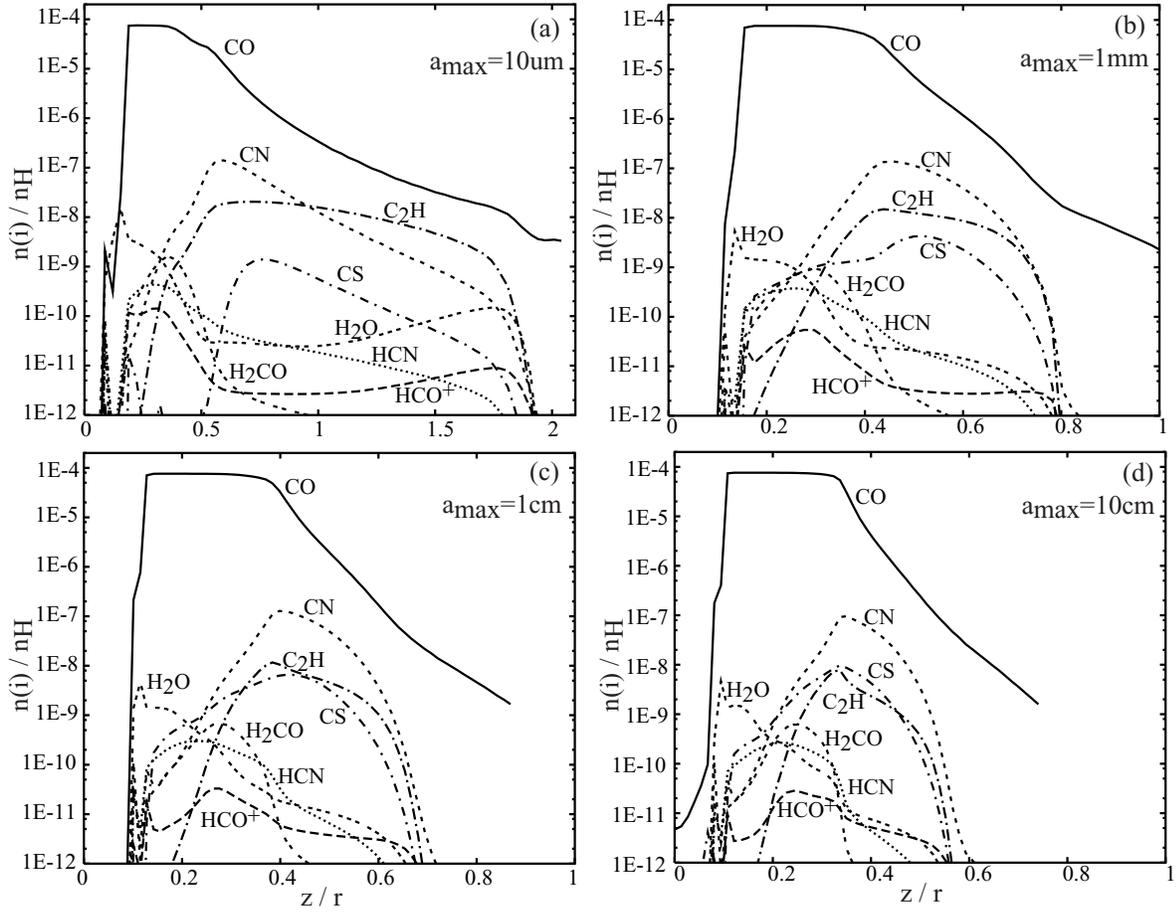}
\caption{Vertical distribution of molecular abundances at a radius of 305 AU
in the models with $a_{\rm max}=10 \mu$m, 1 mm, 1 cm, and 10 cm.
\label{abun_z}}
\end{figure}

Figure \ref{abun_z} $a$ shows the vertical distribution of molecular abundance
at radius of 305 AU in the disk with $a_{\rm max}=10 \mu$m. It is qualitatively
similar to the results of \citet{zah03}. The surface layer is a 
PDR, while heavy element species are depleted onto grains in the midplane
layer. Gaseous molecules are abundant in the intermediate layer, and they
reach their maximum abundance at different heights; CN and C$_2$H reach peak
abundances at larger heights than HCN, H$_2$CO, and H$_2$O. At $z/r\sim 0.1-0.2$
some species show a sharp local peak, which is caused by a selective adsorption
at temperatures of $10-20$ K. Such small structure would, in reality, be smeared out by
turbulent mixing (\S 4.2).

\begin{figure}
\plotone{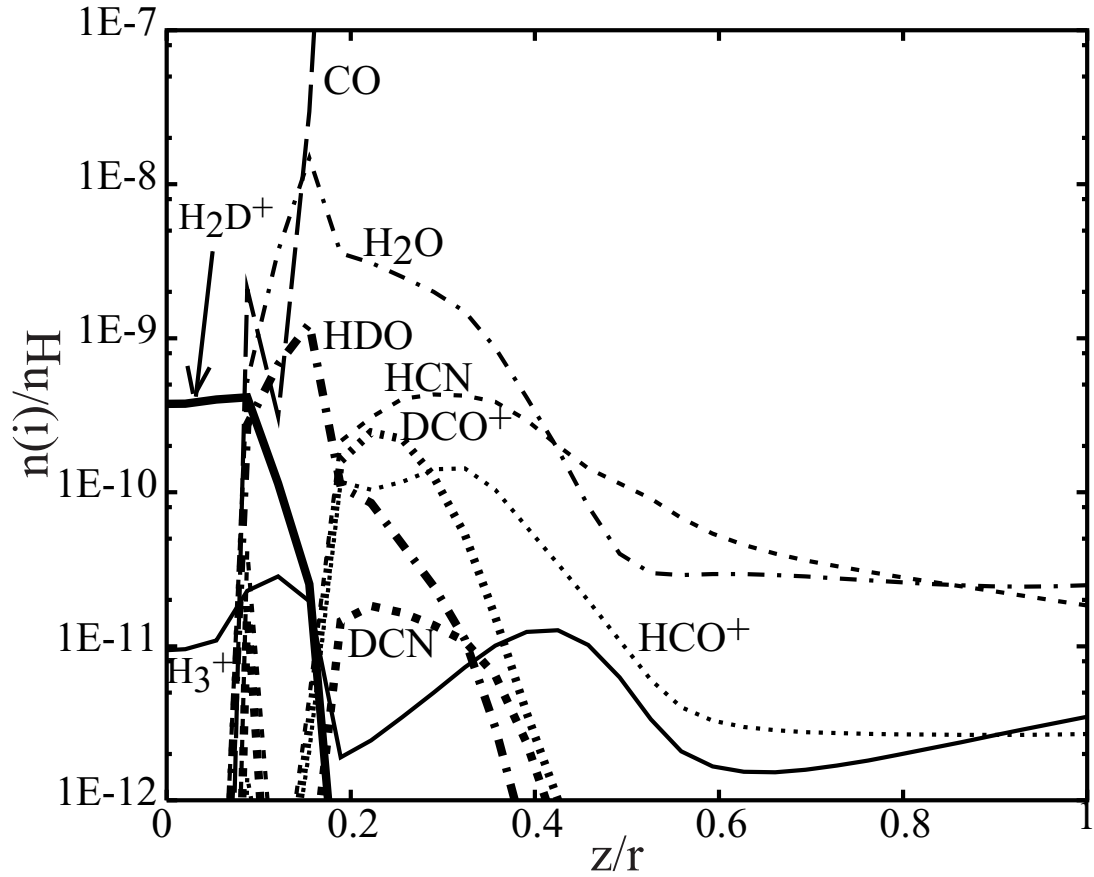}
\caption{Vertical distribution of deuterated species and their normal
counterparts at a radius of 305 AU in the model with $a_{\rm max}=10 \mu$m.
\label{deut}}
\end{figure}

Our reaction network contains mono-deuterated species. Figure \ref{deut}
shows the vertical distribution of some deuterated species and their normal
counterparts at $r=305$ AU in the disk with $a_{\rm max}=10 \mu$m.
Significant deuterium fractionation can be found; heavy element species
such as DCN and HDO are abundant right above the midplane layer,
while H$_2$D$^+$ is more abundant than H$_3^+$ in the cold midplane layer.
Since H$_2$D$^+$ is formed by the reaction H$_3^+$ + HD and destroyed by
reactions with CO and electron, the abundance
ratio of H$_2$D$^+$ to H$_3^+$ becomes very high in the region of heavy CO
depletion and low ionization degree. The abundance ratio of DCO$^+$ to
HCO$^+$ is also higher than unity at $z/r\sim 0.2$. This result, however,
should be taken with caution. It is partly caused by the high abundance of
D atom through D + HCO$^+$ which forms DCO$^+$.
Due to the lack of an approximate method (such as the shielding
factor) to evaluate the photodissociation rate of HD, we assume that its
dependence on UV spectrum is the same as that of CN. If the rate is
evaluated more accurately, i.e., considering the self shielding and mutual
shielding with H$_2$ lines, D atoms and hence DCO$^+$ could be less abundant. 
%In other words, photodissociation of HD could be important for deuterium
%fractionation is disks, and hence should be treated with caution.

\begin{figure}
\plotone{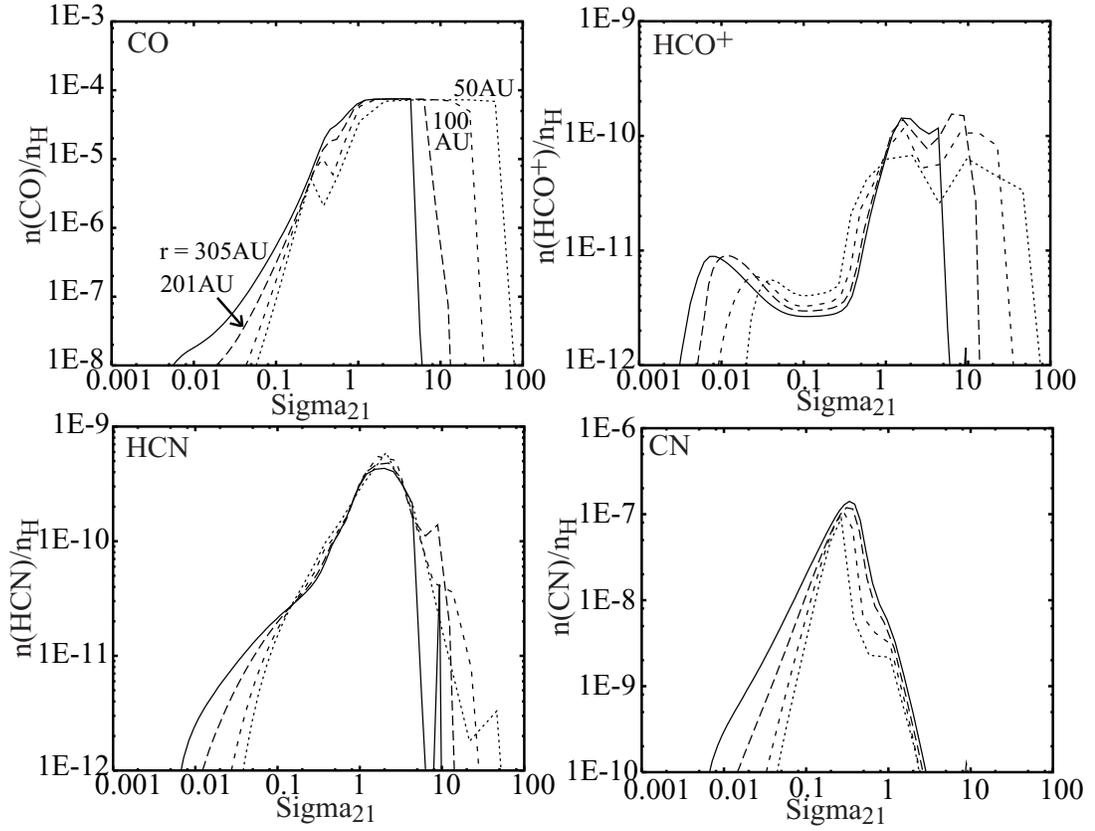}
\caption{Vertical distribution of molecular abundances as a function of
$\Sigma_{21}$ measured from the disk surface at $r=305$ AU (solid), 201 AU
(long-dashed), 100 AU (short-dashed) and 50 AU (dotted line) in the
model with $a_{\rm max}=10 \mu$m.\label{av_abun}}
\end{figure}

Input parameters of the chemical network model are density, temperature and
UV radiation field, which can be approximately represented by $A_{\rm v}$
or the column density of hydrogen nuclei measured from the disk surface
$\Sigma_{\rm H}$.
Since they vary with height ($z$),
it would be interesting to switch the horizontal axis of Figure \ref{abun_z}
to these parameters. In fact, \citet{aik02} described the conditions
for a high abundance of CO and HCN in terms of these parameters.
We checked the dependence of molecular abundances on the three parameters, and
found that for some species the vertical distributions at different
radii are similar, if they are plotted as a function of the hydrogen
column density measured from the disk surface. Figure \ref{av_abun} shows
the abundance of CO, HCO$^+$, HCN and CN at assorted radii as a function of
$\Sigma_{21}\equiv \Sigma_{\rm H}/(1.59\times 10^{21}$ cm$^{-2})$,
the denominator of
which is a conversion factor of the hydrogen column density to $A_{\rm v}$
in the case of interstellar dust.
Although dust grains are larger than the interstellar
dust in our models, we adopt this normalization for convenience.
CO is abundant at $\Sigma_{21}\gtrsim 0.4$,
while HCO$^+$, HCN and CN reach their peak abundance at $\Sigma_{21}\sim 2$,
2 and 0.3, respectively. The lower boundary of the CO and HCO$^+$
layers is determined by the freeze out of CO at $T\sim 20$ K. It is worth
noting that the peak abundances of CO, HCN, and CN are almost independent of
radius. On the other hand, the peak abundance of HCO$^+$ is proportional to
$n_{\rm H}^{-1/2}$ and thus is lower at inner radii, where the layer
of a certain value of $\Sigma_{21}$ has higher density than that at outer
radii. The vertical distribution of molecular abundances obtained in
\citet{zah03} also shows a good correlation with $\Sigma_{21}$
(i.e. $A_{\rm v}$).
Since it is time-consuming to calculate a detailed chemical reaction network
throughout the spatial grid, such a ``scaling law'' would be useful
to make an empirical abundance model for the comparison with observation.

\begin{figure}
\plotone{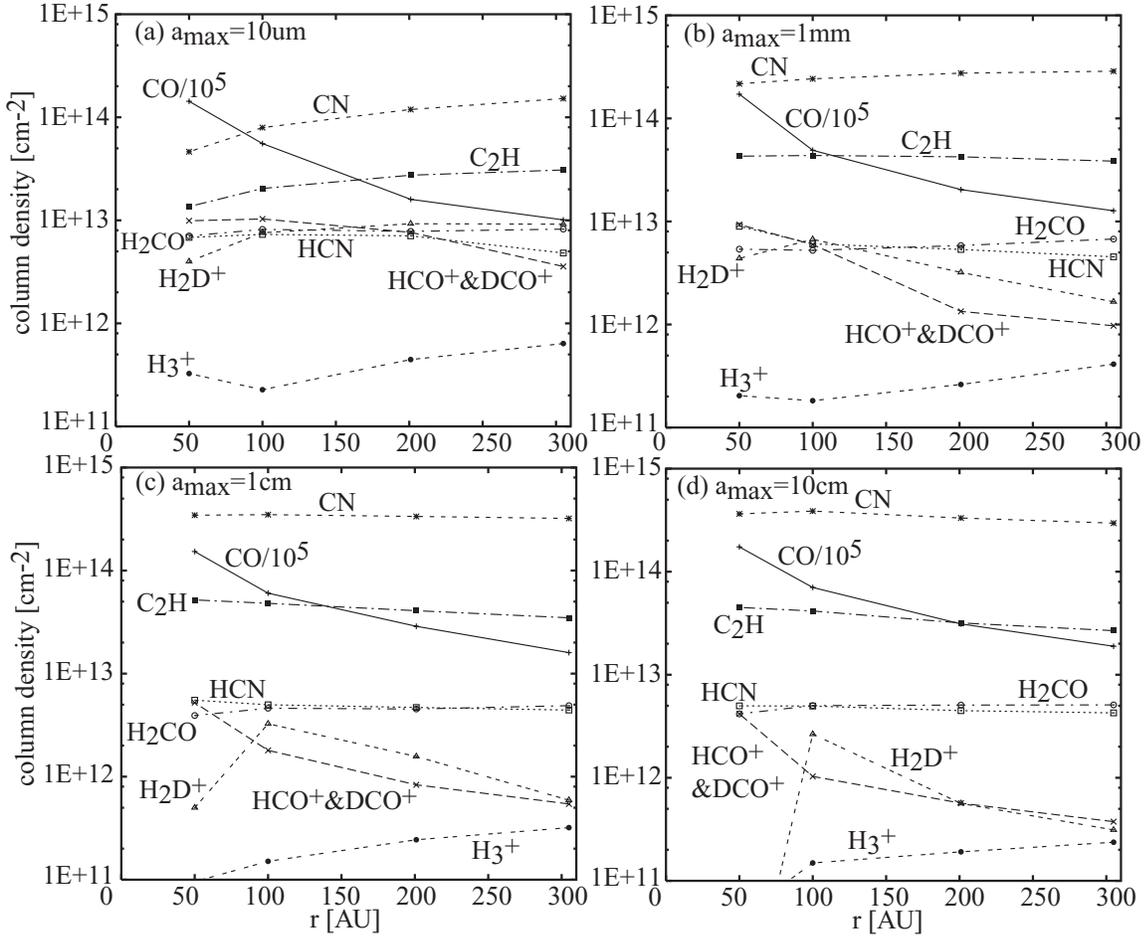}
\caption{Radial distribution of molecular column densities
in the disk models with $a_{\rm max}= 10\mu$m, 1 mm, 1 cm, and 10 cm.
\label{column}}
\end{figure}

Molecular column densities are obtained by integrating the absolute
molecular abundance $n$(i) in the vertical direction. The radial distribution of
assorted molecular column densities in the model with $a_{\rm max}=10 \mu$m is
shown in Figure \ref{column} (a). We plot the sum of HCO$^+$ and DCO$^+$ column
densities, instead of HCO$^+$, because our model may over-produce DCO$^+$ from
HCO$^+$ as stated above.
Carbon monoxide and HCO$^+$ increase towards the center because the amount
of warm gas ($T\gtrsim 20$ K) is larger at the inner radius.
%Column density of HCO$^+$
%also increases towards the center, but the gradient is smaller, because of
%the dependence of its abundance on gas density.
The column densities of H$_2$CO and HCN are almost constant at these radii
($50\lesssim r \lesssim 300$ AU), as expected from the weak dependence of their
vertical distribution on radius (Figure \ref{av_abun}).
A slight decrease of CN and C$_2$H
column densities toward the center is caused by their low abundance in the
layer of $\Sigma_{21}\sim 0.4$ at the inner radius (Figure \ref{av_abun}).
Note that even the CO abundance is low there; a large fraction of carbon is
adsorbed onto grains in the form of C$_5$ or C$_6$. 
These carbon chains are efficiently produced by gas-phase reactions in
such a dense PDR \citep{suz83} and adsorbed onto grains.
At the temperature of $T\sim 20-50$ K, the carbon chains do not sublimate
efficiently, and are selectively accumulated onto grain surfaces
\citep{aik99}. 
The adsorption rates are higher in the inner radius with higher densities.
In the models with larger $a_{\rm max}$ (panel $b-d$ of Figure \ref{column},
see below), column
densities of CN and C$_2$H do not decrease toward the center, because the
time scales for adsorption and selective accumulation of carbon chains are
larger.

\subsubsection{Dependence on Grain Size}
Figure \ref{abun_z} $b-d$ show the vertical distribution of molecular
abundances at $r=305$ AU in the cases of $a_{\rm max}=$ 1mm, 1 cm, and 10 cm.
It should be noted that the range of the horizontal axis of panel $b-d$ is
different from that of panel $a$. The distributions are qualitatively
similar to the model with $a_{\rm max}=10 \mu$m (Figure \ref{abun_z} $a$)
except that they are confined to smaller heights.

If we plot the molecular abundances as a function of $\Sigma_{21}$
instead of $z/r$, they are similar to Figure \ref{av_abun},
except that the peak abundance is reached at larger $\Sigma_{21}$ by a
factor of a few. For example, HCN abundance has a peak at $\Sigma_{21}
\sim 4$ in the model with $a_{\rm max}=10$ cm, compared with  $\Sigma_{21}
\sim 2$ in the $a_{\rm max}=10 \mu$m model. It may sound counter-intuitive,
since the dust opacity at UV wavelength is smaller by 1-2 orders of magnitude
in the large grain models than in the $a_{\rm max}=10 \mu$m model
(Figure \ref{opacity}).
It should be noted that H$_2$ molecule, which is a key ingredient for the
formation of other molecules, is self-shielded from UV radiation, and that
the gas density rapidly increases as a function of depth from the disk surface.
Because of the geometrical effects, it is easier to attenuate the UV radiation
directly coming from the central star than that coming from the vertical
direction (i.e. interstellar radiation and scattering of the stellar
component). Even though UV photons, especially those coming from the vertical
direction, are not much attenuated, rates of two-body reactions
%with H$_2$, and/or atoms
overwhelm the photodissociation rates at a certain depth.
It is worth noting that in the model of \citet{zah03},
which assumed interstellar grains, photodissociation rates are
not necessarily low in the molecular layer (see their Figures 3 and  4).

Consequently, column densities of many molecular species do not much depend
on $a_{\rm max}$ (Figure \ref{column}). Notable exceptions are
HCO$^+$, H$_3^+$ and H$_2$D$^+$. Since the HCO$^+$ abundance is lower at higher
densities, its column density is smaller in the models with larger
$a_{\rm max}$, in which the molecular layer is confined to smaller heights.
Protonated dihydrogen H$_3^+$ and its isotopes are abundant in the cold
midplane, which is thinner in the models with larger $a_{\rm max}$
%(Figure \ref{rhotem_z}).
(Figure \ref{comp_amax}).

%radial dependence on grain size

%\subsection{X-rays}
%X-ray

\section{Discussion}
\subsection{Self consistency of the model}
Chemistry and physical structure in disks are coupled; molecular abundances
depend on physical parameters such as density, and line emission from certain
species such as C$^+$ and O is the main coolant in the surface layers.
It is, however, very time-consuming if we couple the calculation of physical
structure with the detailed time-dependent chemical model. Instead, we have
constructed a physical disk model adopting a chemical equilibrium equation of
\citet{nl97}, and then performed a detailed chemical network
calculation. Here we check the self consistency of our models;
the temperature and density distributions are re-evaluated using
the molecular abundances obtained from the detailed chemical reaction network.

\begin{figure}
\plotone{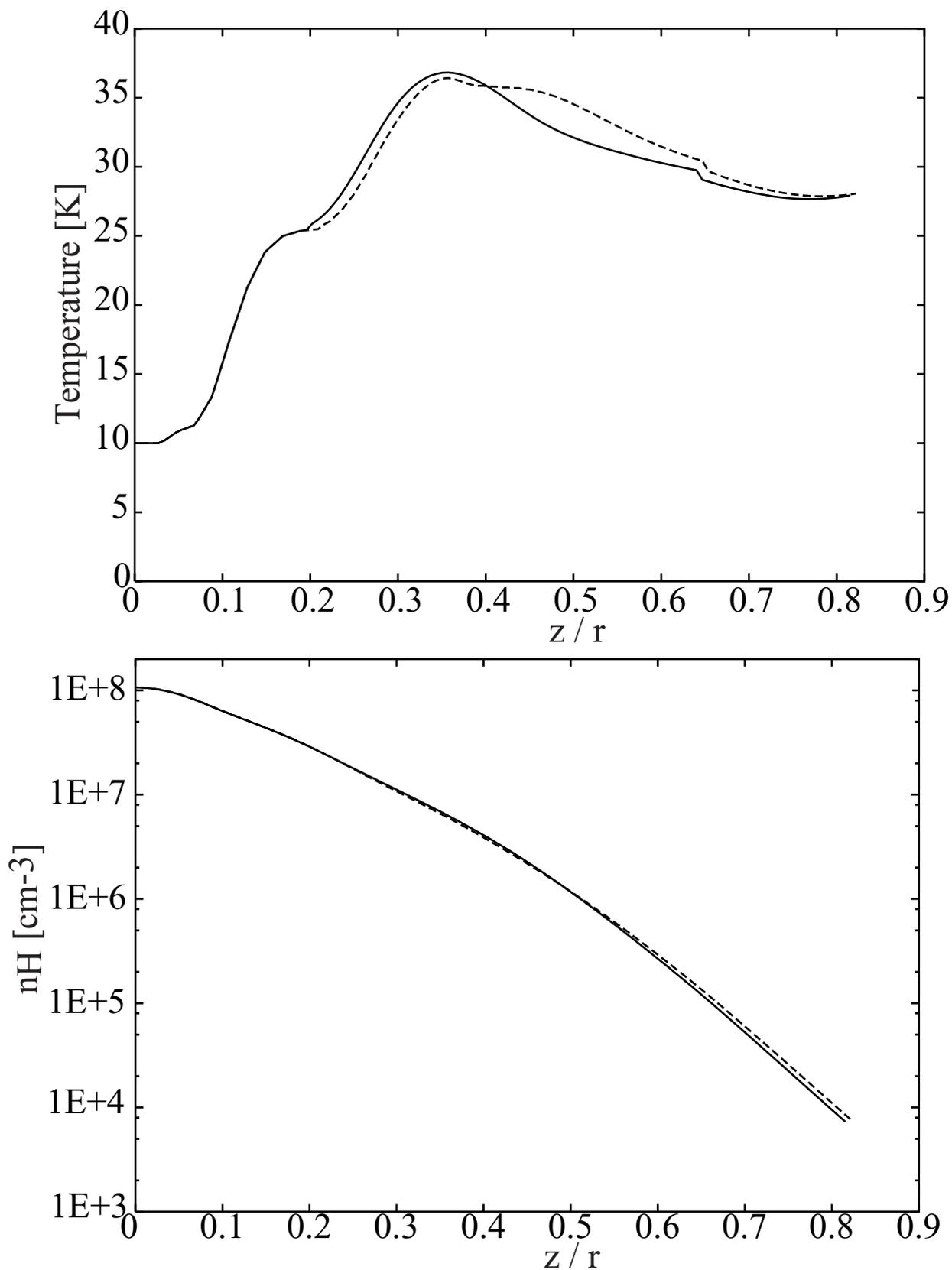}
\caption{Vertical distribution of density and temperature at a radius of
$r=201$ AU and time $t=1\times 10^6$ yr in the model with $a_{\rm max}= 1$ cm.
The solid lines represent our original values, while the dotted lines
are obtained using the molecular abundances from the detailed chemical
reaction network.
\label{re_evaluate}}
\end{figure}

Figure \ref{re_evaluate} shows the vertical distribution of density and
temperature at radius of $r=201$ AU and time $t=1\times 10^6$ yr
in the model with $a_{\rm max}= 1$ cm.
The solid lines represent our original values, while the dotted lines
are obtained using the molecular abundances from the detailed reaction network.
In the surface layer ($z/r\gtrsim 0.4$) the temperature is slightly higher
than the original value; the self shielding of CO, which is not included in the
simple equilibrium equation, lowers the
C$^+$ abundance, which is a major coolant in the surface layer. The gas density
in the surface layer is thus higher than the original value, which then
slightly lowers the gas temperature at $z/r\sim 0.3$.

It should be noted that the difference between the original and re-evaluated
structures is small enough that the detailed chemical reaction network would
give
almost the same abundance distribution for the re-evaluated structure.
Therefore we can conclude that our models of physical structure and
molecular abundances are essentially self-consistent.

\subsection{Effects of hydrodynamical processes}
We did not include hydrodynamic motions such as turbulence within 
the disk.
Such motion will have two major effects on chemistry: (i) it tends to smear the
stratification of molecular abundances and (ii) transport of chemically
active species and/or mother molecules will modify the chemical reaction
network.

The former effect, (i), can be estimated by comparing the mixing and chemical
timescales. Following \citet{aik96}, the mixing timescale over distance $D$
can be written as
\begin{equation}
t_{\rm mix} \approx \frac{D^2}{l_{\rm turb} v_{\rm turb}}.
\label{eq:mix}
\end {equation}
The equation stands for either radial mixing and vertical mixing, as long as
the transport process is turbulence with typical eddy size of $l_{\rm turb}$
and eddy speed of $v_{\rm turb}$.
For example, at radius of 200 AU, the midplane temperature is 10 K, and
hence the sound speed is $\sim 2\times 10^4$ cm s$^{-1}$, and the scale
height is $\sim 36$ AU. If we set $l_{\rm turb}$ and $v_{\rm turb}$ to be
10 \% of the scale height and sound velocity, respectively, it will take
$\sim 9\times 10^4$ yr to mix the matter over the scale height ($\sim 36$ AU).
The chemical timescale is short $\lesssim 10^4$ yr in the surface and
midplane layers of the disk (\S 2.2), while the chemistry in the intermediate
molecular layer is almost in a pseudo-steady state at $10^5-10^6$ yr (\S 2.2).
Therefore the overall three-layer model cannot be smeared out by the mixing
in the vertical direction, although the distribution within the molecular
layer can be somewhat averaged and the molecular layer can be extended
in the vertical direction. For example, \citet{ddg03} observed two rotational
transitions ($J=2-1$ and $1-0$) of $^{13}$CO in DM Tau, and obtained the
gas temperature of 13 K, which is lower than the freeze-out temperature of CO
($\sim 20$ K).
The existence of such cold CO gas can be explained by the vertical mixing.
Equation (\ref{eq:mix}) indicates that at the radius of $\sim 200$ AU CO gas
can migrate over $\sim 10$ AU in the vertical direction within $10^4$ yr,
which is comparable to the freeze-out timescale in the midplane regions
of DM Tau.\footnote{The midplane density in DM Tau is lower than that
in our disk model \citep{dut97}.}

Accretion and radial mixing would not
significantly modify our results, because the vertical distributions of
molecular abundance at different radii are similar at least in the outer
radius (\S 2.2, \S 3.2).

The latter effect, (ii), will enhance the gas-phase abundance of grain-surface
products
in the intermediate molecular layer; species such as H$_2$CO are formed on
the grain surface in the midplane and are sublimated to the gas phase in
the warm molecular layer (Willacy et al. in prep; Semenov et al. in prep).

\subsection{Implications for molecular line observations}
The derivation of molecular abundances from line intensities is not
straightforward because of spatial variation of molecular abundances and
physical condition (i.e. temperature and density). Hence a prediction of
line intensities from model disks is important. Calculation of non-LTE 2D
line transfer, however, is beyond a scope of the present work. Here we make
some qualitative comments on molecular line intensities by comparing the
spatial distribution of molecular abundances and physical conditions.

Firstly, we compare the distributions of gas density (Figure \ref{rhotem_z})
and molecular abundances (Figure \ref{abun_z}). At $z/r\lesssim 0.5$, where
various molecular species reach their peak abundance, the gas density $n_{\rm H}$
is $\gtrsim 10^6$ cm$^{-3}$. Hence most of the major molecular lines in
millimeter wavelength can be excited.
Secondly, distributions of abundances and temperature are compared.
Because of vertical distribution of abundances, each molecular line
and the dust continuum trace different layers of the disk.
Significant amounts of CN, C$_2$H and CO exist in
the upper layer in which gas and dust temperature is not coupled.
As the maximum grain size becomes larger, the gas temperature
in the surface layer decreases, which could be observable through 
these molecular
lines.
Species such as HCN and H$_2$O exist in the layer in which the gas and dust
temperatures are mostly coupled, but are higher than their midplane values.
On the other hand, a major fraction of dust exists in the
midplane. Our results suggest that owing to the vertical temperature gradient
the molecular lines can be observed even in the disks which are optically
thick in the dust continuum.
Although the detailed chemical network is solved only in the outer region
($\ge 50$ AU) of small-mass disks in the present work, the vertical
distribution of temperature and molecular abundances in the surface layers
would be qualitatively similar in the inner regions and/or more massive disks.

%density > 10^6 ... higher than critical density for most lines
%diff temp dust, cn hcn
%line observable even if dust is thick
%dependence on dust size
%line intensity should decrease as dust grows...CO, CN
%               could increase as dust grows ... high critical density lines
%comp temp and abun dist > line obs

\section{Summary}
We have  self-consistently calculated the distribution of density, dust
temperature and gas temperature in protoplanetary disks. The density
is obtained by assuming hydrostatic equilibrium. The dust temperature is
obtained by the 2D radiation transfer, and the gas temperature is determined by
a detailed energy balance among photo-electric heating, line cooling,
and gas-dust heat exchange. Gas is much hotter than dust in the surface layers,
while the gas and dust are energetically well-coupled at smaller heights
because of higher densities.

We have investigated the effect of grain growth on disk structure by varying
the maximum size of dust grains $a_{\rm max}$ in the disk model.
In the models with
larger $a_{\rm max}$, the gas temperature in the surface layers is smaller
because of reduced photo-electric heating rates on small dust particles.
On the other hand, the gas temperature at intermediate heights is higher in the
models with larger $a_{\rm max}$, because of deeper penetration of stellar
radiation. The density distribution is more puffed up in the models with
smaller $a_{\rm max}$.

Molecular abundances at $r=50$, 100, 201, and 305 AU are calculated using a
detailed chemical reaction network. In the surface layer, the dominant chemical
process is photodissociation and/or photoionization by UV radiation from
the central star and interstellar field. At intermediate heights, the gas
density is high, and molecular formation via two-body reactions is more
efficient than photodissociation. The vertical distributions of some
molecular abundance do not depend much on radius, when they are plotted as
a function of the hydrogen column density measured from the disk surface.
%The heights of peak molecular abundance are best described by the
%column density from the disk surface. For example, in the model with
%$a_{\rm max}=10 \mu$m, HCO$^+$, HCN and CN have their peak abundance at
%$\Sigma_{21}\sim 2$, 2 and 0.3, respectively.
%In the midplane of the outer radius $r\gtrsim 50$ AU, heavy element
%species are adsorbed onto grains because of the low temperature and high
%density.
Hence the integrated molecular column densities do not much depend on radius. 
Exceptions are CO and HCO$^+$; they are abundant if the temperature is
higher than the sublimation temperature of CO ($\sim 20$ K), and if CO is
shielded from the UV radiation. The amount of such warm gas is larger at
inner radii.

The dependence of the molecular distributions on grain size ($a_{\rm max}$)
is investigated. In the models with larger $a_{\rm max}$, the geometrical
thickness of the disk is smaller, and the gaseous molecules are confined
to smaller height. However, if we plot
the vertical distribution of molecules as a function of hydrogen column density
from the disk surface, it does not significantly depend on $a_{\rm max}$;
in the model with $a_{\rm max}=10$ cm, the column density of the peak
molecular abundance is larger only by a factor of a few compared with the case
of $a_{\rm max}=10 \mu$m. Hence the integrated molecular column densities
do not much depend on $a_{\rm max}$. Notable exceptions are HCO$^+$, H$_3^+$
and H$_2$D$^+$, which have smaller column densities in the models with
larger $a_{\rm max}$.

Because of the vertical distribution of molecular abundances, layers at
different heights can be traced by choosing appropriate molecular lines.
Significant amounts of CN, C$_2$H and CO exist in
the surface layer in which gas and dust temperatures are not coupled.
Species such as HCN and H$_2$CO exist in the layer in which the gas and dust
temperatures are mostly coupled but are higher than their midplane values,
while a major fraction of dust exists in the midplane.

\acknowledgments

This work is supported by a Grant-in-Aid for Scientific Research (16036205,
17039008) and ``The 21st Century COE Program of Origin and Evolution of
Planetary Systems" of the Ministry of Education, Culture, Sports, Science
and Technology of Japan (MEXT).

\clearpage 

\end{document}